\documentclass{ws-ijmpa}
\usepackage[super,compress]{cite}
\usepackage{graphicx}
\usepackage{braket}
\usepackage{epsfig}
\usepackage{color}

\newcommand{\be}{\begin{equation}}
\newcommand{\ee}{\end{equation}}
\newcommand{\bea}{\begin{eqnarray}}
\newcommand{\eea}{\end{eqnarray}}
\newcommand{\beq}{\begin{equation}}
\newcommand{\eeq}{\end{equation}}
\newcommand{\nn}{\nonumber}

\def\fun#1#2{\lower3.6pt\vbox{\baselineskip0pt\lineskip.9pt
\ialign{$\mathsurround=0pt#1\hfil##\hfil$\crcr#2\crcr\sim\crcr}}}

\begin{document}

\title{
Vertex operators for production of baryon resonances in nucleon-gamma  collisions
with preserving gauge invariance and analyticity}

\author{ A.V. Anisovich, V.V. Anisovich,
A.V. Sarantsev,}

\date{today}
\maketitle

\begin{center}
{\it
National Research Centre ''Kurchatov Institute'':
Petersburg Nuclear Physics Institute, Gatchina, 188300, Russia}

\end{center}

\begin{abstract}

Vertex operators for photo- and electro-production of baryon states
  with arbitrary spin-parity, $ \gamma + N\to B(J^P)$, are constructed.
The operators obey gauge invariance and analyticity constraints.
Analyticity is realized
as a requirement of the generalized Siegert theorem for vertex form
factors.
\end{abstract}
\vspace{0.6 cm}
   PACS numbers: 13.60.Le, 11.55.Jy., 12.40.Yx

\section{Introduction}

The photo- and electro-production of mesons off nucleons
provide a direct way
for the identification of baryon resonances, that was at the times of the
discovery of low-lying states
[\cite{501,502,503,504,gell,kroll,24}] as well as at
recent studies
[\cite{bonn1,bonn2,bonn3,now1,now2,azna,now3,now4,now5}].

The present work is
devoted to the baryon-nucleon-gamma vertex,
$B_{J^P}(p)\to N_{\frac12^+}(k)+\gamma_{1^-}(q)$, where $p,k,q$
are correspondingly four-momenta of baryon state with spin-parity $J^P$, nucleon
and gamma; energy-momentum conservation reads
$p=k+q$ . Our consideration is based on results of ref. [\cite{AS}] where the
case of real photon ($q^2=0$) was considered for  any $J^P$. An advantige of the
used classification of operators is the fixation not only spin-parity but
helicity amplitudes as well that is convenient in data fit procedure.

The case of arbitrary $q^2$ is the subject of this paper.
A problem in the construction of the gauge invariant spin-momentum
operators appears because of the pole singularity ($1/q^2$) inherent to
the spin part of a vector particle propagator. We eliminate the singularity
by adding the longitudinal spin operator, which is nilpotent at $q^2=0$,
  as a cancellation term. Nilpotent operators for improving the analytical
stucture of operators were used for meson states
in [\cite{nilpAAMMNS,nilpASBKNT,nilpAM}]  (see also [\cite{book3,book4}]). The new
operators impose a requirement for the decay form factors which is known for
real photons as the Siegert theorem [\cite{sieg}].

The present consideration is based on spin-operator expansion technique
  presented in detail in
[\cite{book3,book4}], necessary elements of this  technique are given in
Appendices A,B.

\section{Vertices and operators}
According to the classification used in [\cite{AS}]
we have two types of vertices:\\
(i) the $(+)$-vertices for transitions
$B(J^P)\to N(\frac12^+)\gamma$ with spin-parity of the decaying states equal to
$J^P=\frac12^-,\frac32^+,\frac52^-,\frac72^+,...$ and
\\ (ii) the $(-)$-vertices for decaying states with
$J^P=\frac12^+,\frac32^-,\frac52^+,\frac72^-,...$

\subsubsection{The $(+)$-vertices}

Vertices
of the $(+)$-states ($J^P =\frac12^-,\frac32^+,\frac52^-,... $) read:
\bea \label{1}
  &&
G^{(a+)}_{\alpha_1...\alpha_n}(k_\perp) =G^{(a+)}_{n+\frac12}(s, q^2)
V^{(a+)\mu}_{\alpha_1...\alpha_n}(k_\perp)
\epsilon_\mu^{\perp q},\qquad a=1,2,3,
  \\
&&
V^{(1+)\mu}_{\alpha_1...\alpha_n}(k_\perp) =
i\gamma_{\mu}^{\perp p}\gamma_5 X^{(n)}_{\alpha_1...\alpha_n}(k_\perp)
  \,,
\nonumber
\\
&&
V^{(2+)\mu}_{\alpha_1...\alpha_n}(k_\perp) =
i\gamma_\nu^{\perp p}\gamma_5 X^{(n+2)}_{\mu\nu\alpha_1...\alpha_n}(k_\perp)
  \,,
\nonumber
\\
&&
V^{(3+)\mu}_{\alpha_1...\alpha_n}(k_\perp) =
i\gamma_\nu^{\perp p}\gamma_5 X^{(n)}_{\nu\alpha_1...\alpha_{n-1}}(k_\perp)
g_{\alpha_n\mu}^{\perp p}
  \,.
\nonumber
\eea
Here $n=J-\frac12 $ and
\bea
&&
\quad s=p^2=(k+q)^2,\qquad  (qp)=\frac12(p^2-M^2_N+q^2)\\
&&
k=k_\perp +p\frac{(kp)}{p^2},\qquad q=-k_\perp +p\frac{(qp)}{p^2},
\nonumber
\\
&&
\gamma_{\mu}^{\perp p}=
\gamma_{\mu'}g_{\mu'\mu}^{\perp p}=\gamma_{\mu'}
(g_{\mu'\mu}-\frac{p_{\mu'}p_{\mu}}{p^2}).
\nonumber
\eea
The form factors $G^{(a+)}_{n+\frac12}(s, q^2)$ depend on
two momenta squared and the nucleon mass: $s,q^2$ and $k^2=M^2_N$.
Orbital momentum expansion functions, $ X^{(n)}_{\alpha_1...\alpha_n}(k_\perp)$
are presented in Appendix A.
Polarization vectors of gamma obey the constraint for vector particles:
$\epsilon_\mu  q_\mu  =0 $.

\subsubsection{The $(-)$-vertices}

Vertices
of the $(-)$-states ($J^P =\frac12^+,\frac32^-,\frac52^+,... $) read:
\bea \label{3}
  &&
G^{(a-)}_{\alpha_1...\alpha_n}(k_\perp) =G^{(a-)}_{n+\frac12}(s, q^2)
V^{(a-)\mu}_{\alpha_1...\alpha_n}(k_\perp)
\epsilon_\mu^{\perp q},\qquad a=1,2,3
  \\
&&
V^{(1-)\mu}_{\alpha_1...\alpha_n}(k_\perp) =
\gamma_{\nu}^{\perp p}\gamma_\mu
X^{(n+1)}_{\nu\alpha_1...\alpha_n}(k_\perp)
  , \nonumber
\\
&&
V^{(2-)\mu}_{\alpha_1...\alpha_n}(k_\perp) =
   X^{(n+1)}_{\mu\alpha_1...\alpha_n}(k_\perp)
  ,
\nonumber
\\
&&
V^{(3-)\mu}_{\alpha_1...\alpha_n}(k_\perp) =
   X^{(n-1)}_{\alpha_2...\alpha_{L}}(k_\perp)
g_{\alpha_1\mu}^{\perp p}
\nonumber
\eea

\subsection{Operators at $q^2\neq 0$}

Equations (\ref{1}),(\ref{3}) are written for the convolution of operators
with photon
polarization vectors. To activate the photon polarization indices
at $q^2\neq 0$ one can use
the completeness conditions for polarization vectors which are
written as follows:
\be \label{4}
\sum\limits_{a}\epsilon^{\perp q }_{\mu'}(a)
\epsilon^{\perp q*}_\mu (a) =g_{\mu'\mu}-\frac{q_{\mu'}q_\mu}{q^2}
\ee
Therefore we write for operators at $q^2\neq 0$:
\bea \label{5}
&&
G^{(a\pm)\mu}_{\alpha_1...\alpha_n}(k_\perp) =
G^{(a\pm)}_{n+\frac12}(s, q^2)
V^{(a\pm)\mu'}_{\alpha_1...\alpha_n}(k_\perp)
(g_{\mu'\mu}-\frac{q_{\mu'}q_\mu}{q^2}),\qquad a=1,2,3
\eea
The operators (\ref{5}) are appropriate for working
at $q^2\neq 0$, for example, with massive
vector particles, $(q^2=M^2)$.

The pole singularity $(1/q^2)$ can be
eliminated, if it is necessary, by the form factor constraint:
$G^{(a\pm)}_{n+\frac12}(s, q^2=0)=0$.
But in some states it turns out to amusing results.
Another way to eliminate the $(1/q^2)$-pole is related to the introduction
of the longitudinal component.

\section{Longitudinal component and generalized Siegert theorem}

The elimination of the pole singularity can be performed by adding
to the operator $(g_{\mu'\mu}-\frac{q_{\mu'}q_\mu}{q^2})$ a
longitudinal
component (orthogonal to $q$) with an appropriate constant $A$:
\bea \label{6}
&&
\Big(g_{\mu'\mu}-\frac{q_{\mu'}q_\mu}{q^2}
\Big) \to
\bigg[g_{\mu'\mu}-\frac{q_{\mu'}q_\mu}{q^2}+A
q_{\mu'}\Big(q_\mu-\frac{q^2}{(pq)}p_\mu \Big)\bigg]_{A=1/q^2}
\\
&&
\nonumber
=
\bigg[g_{\mu'\mu}-\frac{q_{\mu'}q_\mu}{q^2}+\frac{1}{q^2}
q_{\mu'}\Big(q_\mu-\frac{q^2}{(pq)}p_\mu \Big)\bigg]
=\Big(g_{\mu'\mu}- \frac{q_{\mu'}p_\mu}{(pq)} \Big)
\nonumber
\eea
The singularity $1/q^2$ is cancelled but we face a new singularity at
$(pq)=0$, this one can be eliminated by a zero form factor. But
the particularity consists in the fact that the $1/(pq)$ pole is related
to the longitudinal component, therefore one can construct two transverse
operators without the $1/(pq)$ pole singularities.

Correspondingly, one vertex form factor should satisfy
the generalized Siegert theorem, we nominate this operator (and form
factor) as $a=3$:
\be \label{7}
\bigg[G^{(3\pm)}_{J}(s,q^2)\bigg]_{s=M^2_N-q^2}
=0,\qquad a=3.
\ee
Also, let us emphasize that the longitudinal operator
$L_\mu=(q_\mu-\frac{q^2}{(pq)}p_\mu )$ which is attended at (\ref{6}) is
nilpotent at $q^2=0$: ($L_\mu L_\mu=0$). The nilpotent operators were used
in [\cite{nilpAAMMNS,nilpASBKNT,nilpAM}] for the description of photon
interactions with mesons.

\subsection{Separation of the longitudinal operators and analyticity constraint
}
Separation of the longitudinal operators is performed here in detail.

  \subsubsection{The operators for the $(-)$-states}

To perform the separation of the longitudinal and transverse operators
let us fix the $ V^{G(3-)\mu}_{\alpha_1...\alpha_n}(k_\perp)$ as a basic
one:
\bea\label{8}
V^{G(3-)\mu}_{\alpha_1...\alpha_n}(k_\perp) \equiv
\mathfrak{V}^{(3-)\mu}_{\alpha_1...\alpha_n}(k_\perp)
&=&
   X^{(n-1)}_{\alpha_2...\alpha_{n}}(k_\perp)
g_{\alpha_1\mu'}^{\perp p}
(g_{\mu'\mu}-\frac{q_{\mu'}p_\mu}{(pq)})
\\
& =&
   X^{(n-1)}_{\alpha_2...\alpha_{n}}(k_\perp)
(g_{\alpha_1\mu}-\frac{q_{\alpha_1}p_\mu}{(pq)})\nonumber
\eea
The pole singularity at $(pq)=0$ should be eliminated by requirement
(\ref{7}), that is the generalized Siegert theorem.

Let us now consider the vertex $V^{G(2-)\mu}_{\alpha_1...\alpha_n}(k_\perp) $:
\bea \label{9}
V^{G(2-)\mu}_{\alpha_1...\alpha_n}(k_\perp) &=&
   X^{(n+1)}_{\mu'\alpha_1...\alpha_n}(k_\perp)
(g_{\mu'\mu}-\frac{q_{\mu'}p_\mu}{(pq)})
\\
&=&
   \bigg(
   X^{(n+1)}_{\mu\alpha_1...\alpha_n}(k_\perp) +
   X^{(n+1)}_{\mu'\alpha_1...\alpha_n}(k_\perp)
\frac{k^\perp_{\mu'}p_\mu}{(pq)}\bigg)
  \nonumber
  \\
&=&
   \bigg(
   X^{(n+1)}_{\mu\alpha_1...\alpha_n}(k_\perp) - k_\perp^2\frac{2n-1}{n}
   X^{(n-1)}_{\alpha_2...\alpha_n}(k_\perp)g_{\alpha_1\mu}\bigg.
\nonumber
   \\
\nonumber
   \bigg.
&+& k_\perp^2 \frac{2n-1}{n}
   X^{(n-1)}_{\alpha_2...\alpha_n}(k_\perp)
\Big(g_{\alpha_1\mu}-\frac{q_{\alpha_1}p_\mu}{(pq)}\Big)\bigg).
\nonumber
\eea
The last term is proportional to $V^{G(3-)\mu}_{\alpha_1...\alpha_n}(k_\perp) $
and can be removed.
Recall $k_\perp=-q+p\frac{(pq)}{p^2}$ and  $k_\perp^2=q^2-\frac{(pq)^2}{p^2}$.
Also, we have used:
\be  \label{10}
   X^{(n+1)}_{\mu'\alpha_1...\alpha_n}(k_\perp)
\frac{k^\perp_{\mu'}p_\mu}{(pq)}
= k_\perp^2 \frac{2n-1}{n}
   X^{(n-1)}_{\alpha_2...\alpha_n}(k_\perp)
\Big(-\frac{q_{\alpha_1}p_\mu}{(pq)}\Big).
\ee
We see
that the singular term can be removed by redefinition of the vertex:
  \bea  \label{11}
  V^{G(2-)\mu}_{\alpha_1...\alpha_n}(k_\perp) &\to&
  \mathfrak{V}^{(2-)\mu}_{\alpha_1...\alpha_n}(k_\perp)
  \\
&=& V^{G(2-)\mu}_{\alpha_1...\alpha_n}(k_\perp) -q^2 \frac{2n-1}{n}
V^{G(3-)\mu}_{\alpha_1...\alpha_n}(k_\perp)
\nonumber
\eea
The redefined vertex $\mathfrak{V}^{(2-)\mu}_{\alpha_1...\alpha_n}(k_\perp)$ is
analytical at $(pq)=0$
and $q_\mu\mathfrak{V}^{(2-)\mu}_{\alpha_1...\alpha_n}(k_\perp)=0$.

Redefinition of the $V^{G(1-)\mu}_{\alpha_1...\alpha_n}(k_\perp)$ looks as
follows:
\bea   \label{12}
V^{G(1-)\mu}_{\alpha_1...\alpha_n}(k_\perp)
&=&
\gamma_{\nu}^{\perp p}\gamma_{\mu} X^{(n+1)}_{\nu\alpha_1...\alpha_n}(k_\perp)
-\gamma_{\nu}^{\perp p}\hat q X^{(n+1)}_{\nu\alpha_1...\alpha_n}(k_\perp)
\frac{p_\mu}{(pq)}
\\
&=&
\gamma_{\nu}^{\perp p}\gamma_{\mu} X^{(n+1)}_{\nu\alpha_1...\alpha_n}(k_\perp)
-\gamma_{\nu}^{\perp p}\hat q
k_\nu^\perp\frac{2n+1}{n+1} X^{(n)}_{\alpha_1...\alpha_n}(k_\perp)
\frac{p_\mu}{(pq)}
\nonumber
\\
&=&
\gamma_{\nu}^{\perp p}\gamma_{\mu} X^{(n+1)}_{\nu\alpha_1...\alpha_n}(k_\perp)
- \hat k_\perp \hat q \frac{2n+1}{n+1}
\cdot\frac{2n-1}{n}k_{\alpha_1}^\perp X^{(n)}_{\alpha_2...\alpha_n}(k_\perp)
\frac{p_\mu}{(pq)}
\nonumber
\eea
and
\be
\hat k_\perp \hat q = (pq) - \frac{(pq)}{p^2}\hat p \hat k -
q^2
\ee
To cancel the pole item in (\ref{12}), $\sim 1/(pq)$, we add
$C\mathfrak{V}^{(3-)\mu}_{\alpha_1...\alpha_n}(k_\perp)$
with proper coefficient $C$:
\be \label{13}
C=-q^2\frac{4n^2-1}{n(n+1)}.
\ee
The redefined operator reads:
\be
\mathfrak{V}^{(1-)\mu}_{\alpha_1...\alpha_n}(k_\perp)=
V^{G(1-)\mu}_{\alpha_1...\alpha_n}(k_\perp) - q^2\frac{4n^2-1}{n(n+1)}
V^{G(3-)\mu}_{\alpha_1...\alpha_n}(k_\perp)
\ee

\subsubsection{The operators for the $(+)$-states}

We have three operators:
\bea\label{16}
&&
V^{G(1+)\mu}_{\alpha_1...\alpha_n}(k_\perp) =
i\gamma_{\mu'}^{\perp p}\gamma_5 X^{(n)}_{\alpha_1...\alpha_n}(k_\perp)
(g_{\mu'\mu}-\frac{q_{\mu'}p_\mu}{(pq)})
  \,,
\\
&&
V^{G(2+)\mu}_{\alpha_1...\alpha_n}(k_\perp) =
i\gamma_\nu^{\perp p}\gamma_5 X^{(n+2)}_{\mu'\nu\alpha_1...\alpha_n}(k_\perp)
  (g_{\mu'\mu}-\frac{q_{\mu'}p_\mu}{(pq)})
\,,
\nonumber
\\
&&
V^{G(3+)\mu}_{\alpha_1...\alpha_n}(k_\perp) =
i\gamma_\nu^{\perp p}\gamma_5 X^{(n)}_{\nu\alpha_1...\alpha_{n-1}}(k_\perp)
g_{\alpha_n\mu'}^{\perp p}
(g_{\mu'\mu}-\frac{q_{\mu'}p_\mu}{(pq)})
  \,.
\nonumber
\eea
The reorganized operators
$V^{G(a+)\mu}_{\alpha_1...\alpha_n}(k_\perp) \to
\mathfrak{V}^{(a+)\mu}_{\alpha_1...\alpha_n}(k_\perp)$
(with $a=1,2,3$) read:
\bea
\label{16}
V^{G(1+)\mu}_{\alpha_1...\alpha_n}(k_\perp)\to
\mathfrak{V}^{(1+)\mu}_{\alpha_1...\alpha_n}(k_\perp) &=&
V^{G(1+)\mu}_{\alpha_1...\alpha_n}(k_\perp) -
V^{G(3+)\mu}_{\alpha_1...\alpha_n}(k_\perp)
\\
V^{G(2+)\mu}_{\alpha_1...\alpha_n}(k_\perp)\to
\mathfrak{V}^{(2+)\mu}_{\alpha_1...\alpha_n}(k_\perp)&=&
V^{G(2+)\mu}_{\alpha_1...\alpha_n}(k_\perp) -
q^2 \frac{(2n+1)n}{(2n-1)(n+1)}
V^{G(3+)\mu}_{\alpha_1...\alpha_n}(k_\perp)
\nonumber\\
V^{G(3+)\mu}_{\alpha_1...\alpha_n}(k_\perp)\equiv
\mathfrak{V}^{(3+)\mu}_{\alpha_1...\alpha_n}(k_\perp)
&&
\nonumber\\
=i\gamma_\nu^{\perp p}\gamma_5 \bigg(X^{(n)}_{\nu\alpha_2...\alpha_{n}}(k_\perp)
g_{\alpha_1\mu}^{\perp p}
&-&
  X^{(n)}_{\nu\alpha_2...\alpha_{n}}(k_\perp) q_{\alpha_1}\frac{p_\mu}{(pq)}
  \bigg)
\,.
\nonumber
\eea
We see that all operators are subjects of the gauge requirement:
$q_\mu\mathfrak{V}^{(a\pm)\mu}_{\alpha_1...\alpha_n}(k_\perp)=0$, $a=1,2,3$.

\subsection{The generalized Siegert theorem}

Operators with taken into account form factors read as follows:
\be
\mathfrak{G}^{(a\pm)\mu}_{\alpha_1...\alpha_n}(k_\perp)=
G^{(a\pm)}_{J}(s,q^2)\mathfrak{V}^{(a+)\mu}_{\alpha_1...\alpha_n}(k_\perp),
\qquad a=1,2,3.
\ee
The operators $\mathfrak{V}^{(a+)\mu}_{\alpha_1...\alpha_n}(k_\perp)$
are constricted in such a way that the pole singularity $1/(pq)$ attends at
$(a=3)$-component only. Correspondingly form factors $G^{(3\pm)}(s,q^2)$ obey to
generalized Siegert theorem presented in eq.(\ref{7}):
  $G^{(3\pm)}(s,q^2)\sim (s-M^2_N+q^2)$.

\section{
Different versions for construction of the $(\pm)$-operators}

In the previous Section we constructed $(+)$-operators
($J^P=\frac12^-,\frac32^+,\frac52^-,...$) and
$(-)$-operators ($J^P=\frac12^+,\frac32^-,\frac52^+,...$)
which obey analyticity and gauge invariance constraints that are
$\mathfrak{G}^{(a+)\mu}_{\alpha_1...\alpha_n}(k_\perp)$
and $\mathfrak{G}^{(a-)\mu}_{\alpha_1...\alpha_n}(k_\perp)$.
One can construct other sets of operators,
$\widetilde{\mathfrak{G}}^{(a+)\mu}_{\alpha_1...\alpha_n}(k_\perp)=
\mathfrak{G}^{(a-)\mu}_{\alpha_1...\alpha_n}(k_\perp)\cdot i\gamma_5$
and
$\widetilde{\mathfrak{G}}^{(a-)\mu}_{\alpha_1...\alpha_n}(k_\perp)=
\mathfrak{G}^{(a+)\mu}_{\alpha_1...\alpha_n}(k_\perp)\cdot i\gamma_5$.
\\
  We have four equivalent definitions of the ($\pm$)-operators:
\be \label{13}
\begin{tabular}{l|l}
   $\mathfrak{G}^{(1+)\mu}_{\alpha_1...\alpha_n}(k_\perp)$
&
$\mathfrak{G}^{(1-)\mu}_{\alpha_1...\alpha_n}(k_\perp)$
\\
   $\mathfrak{G}^{(+)\mu}_{\alpha_1...\alpha_n}(k_\perp)$
&
$\mathfrak{G}^{(2-)\mu}_{\alpha_1...\alpha_n}(k_\perp)$
\\
   $\mathfrak{G}^{(3+)\mu}_{\alpha_1...\alpha_n}(k_\perp)$
&
$\mathfrak{G}^{(3-)\mu}_{\alpha_1...\alpha_n}(k_\perp)$
\\
\hline
   $\widetilde{\mathfrak{G}}^{(1+)\mu}_{\alpha_1...\alpha_n}(k_\perp)
=\mathfrak{G}^{(1-)\mu}_{\alpha_1...\alpha_n}(k_\perp)\cdot i\gamma_5$
&
$\widetilde{\mathfrak{G}}^{(1-)\mu}_{\alpha_1...\alpha_n}(k_\perp)
=\mathfrak{G}^{(1+)\mu}_{\alpha_1...\alpha_n}(k_\perp)\cdot i\gamma_5$
\\
   $\widetilde{\mathfrak{G}}^{(2+)\mu}_{\alpha_1...\alpha_n}(k_\perp)
=\mathfrak{G}^{(2-)\mu}_{\alpha_1...\alpha_n}(k_\perp)\cdot i\gamma_5$
&
$\widetilde{\mathfrak{G}}^{(2-)\mu}_{\alpha_1...\alpha_n}(k_\perp)
=\mathfrak{G}^{(2+)\mu}_{\alpha_1...\alpha_n}(k_\perp)\cdot i\gamma_5$
\\
   $\widetilde{\mathfrak{G}}^{(3+)\mu}_{\alpha_1...\alpha_n}(k_\perp)
=\mathfrak{G}^{(3-)\mu}_{\alpha_1...\alpha_n}(k_\perp)\cdot i\gamma_5$
&
$\widetilde{\mathfrak{G}}^{(3-)\mu}_{\alpha_1...\alpha_n}(k_\perp)=
\mathfrak{G}^{(3+)\mu}_{\alpha_1...\alpha_n}(k_\perp)\cdot i\gamma_5$
  \end{tabular}
\ee
Obviously, only the $(3\pm)$-operators obey the Siegert theorem requirement.

\section{Conclusion}
We construct gauge invariant operators for vertices  of baryon resonances
decaying into the nucleon-gamma state, $B(J^P)\to N(\frac12^+)+\gamma$.
The operators describe transition vertices of resonances with arbitrary
spin-parity, $J^P$, and arbitrary momenta squared of gamma, $q^2$. The
suggested operator construction presents the generalization of the scheme
of ref. [\cite{AS}] for photon vertices, ($q^2=0$).

The key point in the construction of the gauge invariant operators valid
at all possible $q^2$ is the introduction of longitudinal spin
operator $L_\mu$ which is nilpotent at $q^2=0$: $[L_\mu L_\mu]_{q^2=0}=0$.
The addition of the nilpotent operator corrects the analytical structure
of spin operators.
For meson states such a correction was carried out in
[\cite{nilpAAMMNS,nilpASBKNT,nilpAM}]
(see also [\cite{book3,book4}]). The procedure
for baryon-gamma states is performed here (Section 3): the pole singularity
inherent to spin operators of vector particles ($1/q^2$) is eliminated by
adding the nilpotent operator $L_\mu $.

Form factors of the transitions $B(J^P)\to N(\frac12^+)+\gamma$ should
satisfy
the generalized Siegert requirement:
   $G^{(3\pm)}_{J}(s,q^2)=0$ at $s=M^2_N-q^2$.

\subsection*{Acknowledgement}
   We thank M.A. Matveev, V.A. Nikonov, J. Nyiri for useful discussions.
The paper was supported by grant RSF 16-12-10267.

\section* {Appendix A:  Angular momentum operators for two-meson systems}

We use angular momentum operators
$X^{(L)}_{\mu_1\ldots\mu_L}(k^\perp)$,
$\,Z^{\alpha}_{\mu_1\ldots\mu_L}(k^\perp)$ and the projection operator
$O^{\mu_1\ldots\mu_L}_{\nu_1\ldots\nu_L}(\perp P)$ (see
[\cite{book3,book4,nilpAAMMNS}]).
Let
us recall their definition.

The operators are constructed from the relative momenta
$k^\perp_\mu$ and tensor $g^\perp_{\mu\nu}$. Both of them are
orthogonal to the total momentum of the system:
\be
k^\perp_\mu=\frac12 g^\perp_{\mu\nu}(k_1-k_2)_\nu =k_{1\nu} g^{\perp
P}_{\nu\mu} =-k_{2\nu} g^{\perp P}_{\nu\mu} ,  \qquad
g^\perp_{\mu\nu}=g_{\mu\nu}-\frac{P_\mu P_\nu}{s}\;.
\ee

The operator for $L=0$ is a scalar (we write $X^{(0)}(k^\perp)=1$),
and the operator for $L=1$ is a vector, $X^{(1)}_\mu=k^\perp_\mu $.
The operators $X^{(L)}_{\mu_1\ldots\mu_L}$ for $L\ge 1$ can be written
in the form of a recurrency relation:
\bea
X^{(L)}_{\mu_1\ldots\mu_L}(k^\perp)&=&k^\perp_\alpha
Z^{\alpha}_{\mu_1\ldots\mu_L}(k^\perp)\equiv  k^\perp_\alpha
Z_{\mu_1\ldots\mu_L,\alpha}(k^\perp)  ,
\nonumber\\
Z^{\alpha}_{\mu_1\ldots\mu_L}(k^\perp)&\equiv &
Z_{\mu_1\ldots\mu_L,\alpha}(k^\perp)=
\frac{2L-1}{L^2}\Big (
\sum^L_{i=1}X^{{(L-1)}}_{\mu_1\ldots\mu_{i-1}\mu_{i+1}\ldots\mu_L}(k^\perp)
g^\perp_{\mu_i\alpha}-
\nonumber \\
      -\frac{2}{2L-1}  \sum^L_{i,j=1 \atop i<j}
&g^\perp_{\mu_i\mu_j}&
X^{{(L-1)}}_{\mu_1\ldots\mu_{i-1}\mu_{i+1}\ldots\mu_{j-1}\mu_{j+1}
\ldots\mu_L\alpha}(k^\perp) \Big ).
\label{Vz}
\eea
      We have a convolution equality
$X^{(L)}_{\mu_1\ldots\mu_{L}}(k^\perp)k^\perp_{\mu_L}=k^2_\perp
X^{(L-1)}_{\mu_1\ldots\mu_{L-1}}(k^\perp)$, with $k^2_\perp\equiv
k^\perp_{\mu}k^\perp_{\mu}$, and the tracelessness property of
$X^{(L)}_{\mu\mu\mu_3\ldots\mu_{L}}=0$. On this basis, one can write
down the normalization condition for orbital angular operators:
\bea
\int\frac{d\Omega}{4\pi}
X^{(L)}_{\mu_1\ldots\mu_{L}}(k^\perp)X^{(L)}_{\mu_1\ldots\mu_{L}}
(k^\perp)
      = \alpha_L k^{2L}_\perp \; ,\quad
\alpha_L\ =\ \prod^L_{l=1}\frac{2l-1}{l}  ,
\label{Valpha}
\eea
      where the integration is performed over spherical variables
$\int d\Omega/(4\pi)=1$.

Iterating Eq. (\ref{Vz}), one obtains the
following expression for the operator $X^{(L)}_{\mu_1\ldots\mu_L}$
at $L\ge 1$:
\bea
\label{Vx-direct}
&&X^{(L)}_{\mu_1\ldots\mu_L}(k^\perp)=
\alpha_L \bigg [
k^\perp_{\mu_1}k^\perp_{\mu_2}k^\perp_{\mu_3}k^\perp_{\mu_4}
\ldots k^\perp_{\mu_L}- \\
&&-\frac{k^2_\perp}{2L-1}\bigg(
g^\perp_{\mu_1\mu_2}k^\perp_{\mu_3}k^\perp_{\mu_4}\ldots
k^\perp_{\mu_L}
+g^\perp_{\mu_1\mu_3}k^\perp_{\mu_2}k^\perp_{\mu_4}\ldots
k^\perp_{\mu_L} + \ldots \bigg)+
\nonumber \\
&&+\frac{k^4_\perp}{(2L\!-\!1)(2L\!-\!3)}\bigg(
g^\perp_{\mu_1\mu_2}g^\perp_{\mu_3\mu_4}k^\perp_{\mu_5}
k^\perp_{\mu_6}\ldots k^\perp_{\mu_L}
\nn \\
&&+
g^\perp_{\mu_1\mu_2}g^\perp_{\mu_3\mu_5}k^\perp_{\mu_4}
k^\perp_{\mu_6}\ldots k^\perp_{\mu_L}+
\ldots\bigg)+\ldots\bigg ]. \nonumber
\eea
For the projection operators, one has:
\bea
&&\hspace{-6mm}
O= 1 ,\qquad
O^\mu_\nu (\perp P)=g_{\mu\nu}^\perp \, ,
\nonumber \\
&&\hspace{-6mm}O^{\mu_1\mu_2}_{\nu_1\nu_2}(\perp P)=
\frac 12 \left (
g_{\mu_1\nu_1}^\perp  g_{\mu_2\nu_2}^\perp \!+\!
g_{\mu_1\nu_2}^\perp  g_{\mu_2\nu_1}^\perp  \!- \!\frac 23
g_{\mu_1\mu_2}^\perp  g_{\nu_1\nu_2}^\perp \right ).
\eea
For higher states, the operator can be calculated using the
recurrent expression:
\bea
&&O^{\mu_1\ldots\mu_L}_{\nu_1\ldots\nu_L}(\perp P)=
\frac{1}{L^2} \bigg (
\sum\limits_{i,j=1}^{L}g^\perp_{\mu_i\nu_j}
O^{\mu_1\ldots\mu_{i-1}\mu_{i+1}\ldots\mu_L}_{\nu_1\ldots
\nu_{j-1}\nu_{j+1}\ldots\nu_L}(\perp P)- \nonumber
      \\
&&- \frac{4}{(2L-1)(2L-3)} \times \sum\limits_{i<j\atop k<m}^{L}
g^\perp_{\mu_i\mu_j}g^\perp_{\nu_k\nu_m}
O^{\mu_1\ldots\mu_{i-1}\mu_{i+1}\ldots\mu_{j-1}\mu_{j+1}\ldots\mu_L}_
{\nu_1\ldots\nu_{k-1}\nu_{k+1}\ldots\nu_{m-1}\nu_{m+1}\ldots\nu_L}(\perp
P) \bigg ).
\eea
The projection operators obey the relations:
\bea
O^{\mu_1\ldots\mu_L}_{\nu_1\ldots\nu_L}(\perp P)
X^{(L)}_{\nu_1\ldots\nu_L}(k^\perp)&=&
X^{(L)}_{\mu_1\ldots\mu_L}(k^\perp)\, ,\nonumber \\
O^{\mu_1\ldots\mu_L}_{\nu_1\ldots\nu_L}(\perp P) k_{\nu_1} k_{\nu_2}
\ldots k_{\nu_L} &=& \frac
{1}{\alpha_L}X^{(L)}_{\mu_1\ldots\mu_L}(k^\perp) .
\eea
      Hence, the product of the two $X^L(k_\perp)$ operators results in the
Legendre polynomials as follows:
\be
X^{(L)}_{\mu_1\ldots\mu_L}(k_1^\perp) (-1)^L
O^{\mu_1\ldots\mu_L}_{\nu_1\ldots\nu_L}(\perp P)
X^{(L)}_{\nu_1\ldots\nu_L}(k_2^\perp)\!=\!\alpha_L
\Big(\sqrt{-k_1^{\perp 2}}\sqrt{-k_2^{\perp 2}}\Big)^L P_L(z),
\ee
where $z\equiv (-{ k}_{1\nu}^\perp { k}_{2\nu}^\perp)/(
\sqrt{-k_1^{\perp 2}}\sqrt{-k_2^{\perp 2}})$.

\section*{Appendix B:  Projection operators for baryon resonance states
with arbitrary $J$.}

The wave function of a resonance state with spin $J=n+1/2$, momentum
   $p$ and the effective mass term $\sqrt{s}$ is given by a tensor
four-spinor $\psi_{\mu_1\ldots\mu_n}$. It satisfies the constraints
\be
\label{5-8} (\hat p-\sqrt{s})\psi_{\mu_1\ldots\mu_n}=0, \quad
p_{\mu_i}\psi_{\mu_1\ldots\mu_n}=0,\quad
\gamma_{\mu_i}\psi_{\mu_1\ldots\mu_n}=0,
\ee
    and the symmetry properties
    \bea \label{5-9}
\psi_{\mu_1\ldots\mu_i\ldots\mu_j\ldots\mu_n}&=&
\psi_{\mu_1\ldots\mu_j\ldots\mu_i\ldots\mu_n}\;,
\nonumber \\
g_{\mu_i\mu_j}\psi_{\mu_1\ldots\mu_i\ldots\mu_j\ldots\mu_n}&=&
g^{\perp p}
_{\mu_i\mu_j}\psi_{\mu_1\ldots\mu_i\ldots\mu_j\ldots\mu_n}=0 .
\eea
Conditions (\ref{5-8}), (\ref{5-9}) define the structure of the
denominator of the fermion propagator (the projection operator)
which can be written in the following form:
\bea \label{5-10}
&&
F^{\mu_1\ldots\mu_n}_{\nu_1\ldots \nu_n}(p)=(-1)^n
(\hat p+\sqrt{s}) \Phi^{\mu_1\ldots\mu_n}_{\nu_1\ldots \nu_n}(\perp p),\\
&&
\mathfrak{P}^{\mu_1\ldots\mu_n}_{\nu_1\ldots \nu_n}(p)=
\frac{F^{\mu_1\ldots\mu_n}_{\nu_1\ldots \nu_n}(p)}{m^2-s-i\,m\Gamma(s)}
\nonumber\eea
The operator $\Phi^{\mu_1\ldots\mu_n}_{\nu_1\ldots
\nu_n}(\perp p)$ describes the tensor structure of the propagator.
It is equal to 1 for a ($J=1/2$)-particle and is proportional to
$g^{\perp p}_{\mu\nu}-\gamma^\perp_\mu\gamma^\perp_\nu/3$ for a
particle with spin $J=3/2$ (remind that $\gamma^\perp_\mu=g^{\perp
p}_{\mu\nu}\gamma_\nu$).

The conditions (\ref{5-9}) are identical for fermion and boson
projection operators and therefore the fermion projection operator
can be written as:
\be  \label{5-11}
\Phi^{\mu_1\ldots\mu_n}_{\nu_1\ldots \nu_n}(\perp p)=
O^{\mu_1\ldots\mu_n}_{\alpha_1\ldots \alpha_n} (\perp p)
\phi^{\alpha_1\ldots\alpha_n}_{\beta_1\ldots \beta_n}(\perp p)
O^{\beta_1\ldots \beta_n}_{\nu_1\ldots\nu_n} (\perp p)\ .
\ee
    The operator $\phi^{\alpha_1\ldots\alpha_n}_{\beta_1\ldots \beta_n}
(\perp p)$ can be expressed in a rather simple form since all
symmetry and orthogonality conditions are imposed by $O$-operators.
First, the $\phi$-operator is constructed of metric tensors only, which
act in the space of $\perp p$ and $\gamma^\perp$-matrices. Second, a
construction like $ \gamma^\perp_{\alpha_i}\gamma^\perp_{\alpha_j}=
\frac12 g^\perp_{\alpha_i\alpha_j}+\sigma^\perp_{\alpha_i\alpha_j}$
(remind that here $\sigma^\perp_{\alpha_i\alpha_j}=\frac 12
(\gamma^\perp_{\alpha_i}\gamma^\perp_{\alpha_j}-
\gamma^\perp_{\alpha_j}\gamma^\perp_{\alpha_i}$) gives zero if
multiplied by an $O^{\mu_1\ldots\mu_n}_{\alpha_1\ldots
\alpha_n}$-operator: the first term is due to the traceless
conditions and the second one to symmetry properties. The only
structures which can then be constructed are
$g^\perp_{\alpha_i\beta_j}$ and $\sigma^\perp_{\alpha_i\beta_j}$.
Moreover, taking into account the symmetry properties of the
$O$-operators,  one can use any pair of indices from sets
$\alpha_1\ldots\alpha_n$ and $\beta_1\ldots \beta_n$, for example,
$\alpha_i\to \alpha_1$ and $\beta_j\to \beta_1$. Then
\be
\phi^{\alpha_1\ldots\alpha_n}_{\beta_1\ldots \beta_n}(\perp p)=
\frac{n+1}{2n\!+\!1} \big( g^\perp_{\alpha_1\beta_1}-
\frac{n}{n\!+\!1}\sigma^\perp_{\alpha_1\beta_1} \big)
\prod\limits_{i=2}^{n}g^\perp_{\alpha_i\beta_i} .
\label{5-t1}
\ee
    Since $\Phi^{\mu_1\ldots\mu_n}_{\nu_1\ldots \nu_n}(\perp p)$ is
determined by convolutions of $O$-operators, see Eq. (\ref{5-11}),
we can replace in (\ref{5-11})
\be \label{5-13}
\hspace{-7mm}
\phi^{\alpha_1\ldots\alpha_n}_{\beta_1\ldots \beta_n}(\perp p) \to
\phi^{\alpha_1\ldots\alpha_n}_{\beta_1\ldots \beta_n}( p) =
\frac{n+1}{2n\!+\!1} \big( g_{\alpha_1\beta_1}-
\frac{n}{n\!+\!1}\sigma_{\alpha_1\beta_1} \big)
\prod\limits_{i=2}^{n}g_{\alpha_i\beta_i} .
\ee
    The coefficients in (\ref{5-13}) are chosen to satisfy the constraints
(\ref{5-8}) and the convolution condition:
\be  \label{5-14}
F^{\mu_1\ldots\mu_n}_{\alpha_1\ldots \alpha_n}( p) F^{\alpha_1\ldots
\alpha_n}_{\nu_1\ldots \nu_n}( p)=(-1)^n
F^{\mu_1\ldots\mu_n}_{\nu_1\ldots \nu_n}( p)\cdot 2\sqrt s \;.
\ee

The amplitude of photo-production (or, electro-production) of baryon resonances on
nucleon target reads:
\be
\mathfrak{A}_{\mu_1\ldots\mu_n}(p)=
\frac{F^{\mu_1\ldots\mu_n}_{\alpha_1\ldots \alpha_n}(p)}{m^2-s-i\,m\Gamma(s)}
\mathfrak{G}^{(a\pm)}_{\alpha_1...\alpha_n}(k_\perp)\,, \ee
remind, $a=1,2,3$.

\end{document}